\documentclass[12pt]{article}
\usepackage{epsfig}
\usepackage{amssymb}
\usepackage{amsmath}
\usepackage{amsfonts}
\usepackage{graphicx}
\usepackage{mathrsfs}
\usepackage[dvips]{color}
\usepackage{multirow}

\newcommand{\bsigma}{\boldsymbol{\sigma}}

\newcommand{\R}{\mathbb{R}}

\newcommand{\fa}{\mathfrak{a}}
\newcommand{\fb}{\mathfrak{b}}

\newcommand{\fK}{\mathfrak{K}}

\newcommand{\bfe}{\mathbf{e}}

\newcommand{\bm}{{\mathbf{m}}}

\newcommand{\bzero}{\mathbf{0}}

\newcommand{\bH}{\mathbf{H}}
\newcommand{\bI}{\mathbf{I}}

\newcommand{\bM}{\mathbf{M}}

\newcommand{\bS}{\mathbf{S}}

\newcommand{\cA}{{\mathcal{A}}}

\newcommand{\cH}{\mathcal{H}}

\newcommand{\cK}{\mathcal{K}}

\newcommand{\cP}{\mathcal{P}}

\newcommand{\cR}{\mathcal{R}}
\newcommand{\cS}{\mathcal{S}}
\newcommand{\cT}{\mathcal{T}}
\newcommand{\cU}{\mathcal{U}}

\newcommand{\be}{\begin{equation}}
\newcommand{\ee}{\end{equation}}
\newcommand{\bea}{\begin{eqnarray}}
\newcommand{\eea}{\end{eqnarray}}
\newcommand{\nn}{\nonumber}

\newcommand{\ed}{\end{document}}

\newcommand{\bi}{\begin{itemize}}
\newcommand{\ei}{\end{itemize}}

\newcommand{\bce}{\begin{center}}
\newcommand{\ece}{\end{center}}

\newcommand{\sS}{\mathscr{S}}

\newcommand{\RE}{{\rm Re}}
\newcommand{\IM}{{\rm Im}}

\newcommand{\bcK}{{\boldsymbol{\cK}}}

\newcommand{\bcH}{{\boldsymbol{\cH}}}

\newcommand{\bcU}{{\boldsymbol{\cU}}}

\newcommand{\bPsi}{{\boldsymbol{\Psi}}}

\newcommand{\for}{{\rm for}}

\begin{document}

\title{Low-frequency scattering defined by the Helmholtz equation in one dimension}

\author{Farhang Loran\thanks{E-mail address: loran@iut.ac.ir}~ and
Ali~Mostafazadeh\thanks{E-mail address:
amostafazadeh@ku.edu.tr}\\[6pt]
$^{*}$Department of Physics, Isfahan University of Technology, \\ Isfahan 84156-83111, Iran\\[6pt]
$^\dagger$Departments of Mathematics and Physics, Ko\c{c}
University,\\  34450 Sar{\i}yer, Istanbul, Turkey}

\date{ }
\maketitle

\begin{abstract}

The Helmholtz equation in one dimension, which describes the propagation of electromagnetic waves in effectively one-dimensional systems, is equivalent to the  time-independent Schr\"odinger equation. The fact that the potential term entering the latter is energy-dependent obstructs the application of the results on low-energy quantum scattering in the study of the low-frequency waves satisfying the Helmholtz equation. We use a recently developed dynamical formulation of stationary scattering to offer a comprehensive treatment of the low-frequency scattering of these waves for a general finite-range scatterer. In particular, we give explicit formulas for the coefficients of the low-frequency series expansion of the transfer matrix of the system which in turn allow for determining the low-frequency expansions of its reflection, transmission, and absorption coefficients. Our general results reveal a number of interesting physical aspects of low-frequency scattering particularly in relation to permittivity profiles having balanced gain and loss.
\vspace{2mm}


\end{abstract}

\section{Introduction}

Low-frequency scattering of electromagnetic waves by conductors or dielectric material confined to compact regions of space is a century-old subject of great attention among mathematicians, physicists, and engineers. The subject has an extremely wide range of scientific and technological applications, and there are numerous research articles and monographs covering its various aspects \cite{newton-book,dassios-book}.  The research done by generations of scientists on low-frequency electromagnetic waves makes use of the theory of partial differential equations and leads to implicit series expansions for the scattering data whose evaluation often requires elaborate numerical treatments. This is in contrast to the status of low-frequency (often called low-energy) quantum scattering theory, which is defined through the stationary Schr\"odinger equation. At least for finite-range and exponentially decaying potentials in one dimension, there are iterative schemes allowing for an analytic calculation of the coefficients of the low-frequency series expansion of the scattering data \cite{bolle-1985,bolle-1987,p164}. This observation motivates the use of similar approaches for dealing with the scattering of the low-frequency electromagnetic waves in effectively one-dimensional systems.  

Consider the scattering of a time-harmonic electromagnetic wave with angular frequency $\omega$ by an infinite non-magnetic dielectric slab $S$ with planar symmetry and thickness $\ell$ that is placed in vacuum. We can choose a Cartesian coordinate system $\{(x,y,z)\}$ in which the complex permittivity profile of this system takes the form, $\varepsilon(x;k):=\varepsilon_0\hat\varepsilon(x;k)$, where $k:=\omega/c$ is the wavenumber, $\varepsilon_0$ and $c$ are respectively the permittivity and speed of light in vacuum, 
	\be
	\hat\varepsilon(x;k):=\left\{\begin{array}{ccc}
	w(x/\ell;k) &\for & x\in [a,a+\ell],\\
	1 &\for & x\notin [a,a+\ell],\end{array}\right.
	\label{e1}
	\ee
is the relative permittivity of the system, $w$ is a complex-valued function of $\hat x:=x/\ell$ and $k$ which describes the inhomogeneity of the slab and its dispersion, and $a$ is a real parameter specifying the location of the slab. Throughout this article we assume that  $w$ is bounded, i.e., there is a positive real number $b$ such that for all $\hat x\in[\frac{a}{\ell},\frac{a}{\ell}+1]$ and $k\in\R$, 
		\be 
		|w(\hat x;k)|\leq b^2.
		\label{bound-zero}
		\ee

If the wave is transverse electric (TE) and propagates along the $x$-axis, we can align the $y$- and $z$-axes in such a way that the electric field takes the form $E_0 e^{-i\omega t}\psi(x;k)\bfe_y$. Here $E_0$ is a constant amplitude, $\psi(x;k)$ is a solution of the Helmholtz equation,
	\be
	\partial_x^2\psi(x;k)+k^2\hat\varepsilon(x;k)\psi(x;k)=0,
	\label{H-eq}
	\ee
and $\bfe_y$ is the unit vector pointing along the positive $y$-axis. We can identify (\ref{H-eq}) with the time-independent Schr\"odinger equation, 
	\be
	-\partial_x^2\psi(x;k)+v(x;k)\psi(x;k)=k^2\psi(x;k),
	\label{sch-eq}
	\ee
for the finite-range potential,
	\be
	v(x;k):=k^2[1-\hat\varepsilon(x;k)]=\left\{\begin{array}{ccc}
	k^2[1-w(x/\ell;k)] &\for & x\in [a,a+\ell],\\
	0 &\for & x\notin [a,a+\ell],\end{array}\right.
	\label{v=}
	\ee
which is energy-dependent. The latter feature does not, however, cause any difficulty in employing the standard approach to potential scattering in the study of the scattering problem defined by (\ref{sch-eq}). 

According to (\ref{e1}), there are complex-valued coefficient functions $A_\pm(k)$ and $B_\pm(k)$ such that every solution of the Helmholtz equation  (\ref{H-eq}) satisfies
	\be
	\psi(x;k)=\left\{\begin{array}{ccc}
	A_-(k)e^{ikx}+B_-(k)e^{-ikx} & \for & x\leq a,\\
	A_+(k)e^{ikx}+B_+(k)e^{-ikx} & \for & x\geq a+\ell.
	\end{array}\right.
	\label{asym}
	\ee
The $2\times 2$ matrix $\bM(k)$ that relates $A_\pm(k)$ and $B_\pm(k)$  according to
	\be
	\left[\begin{array}{c}
	A_+(k)\\
	B_+(k)\end{array}\right]=\bM(k)\left[\begin{array}{c}
	A_-(k)\\
	B_-(k)\end{array}\right],
	\label{M-def}
	\ee
is called the transfer matrix of the system \cite{jones-1941,abeles,thompson,yeh,pereyra,griffiths,sanchez,tjp-2020}.
	
Among the solutions of (\ref{H-eq}), there are the left/right-incident scattering solutions $\psi_{l/r}(x;k)$. These satisfy
	\bea
	\psi_l(x;k)&=&\left\{\begin{array}{ccc}
	e^{ikx}+R^l(k)e^{-ikx} & \for & x\leq a,\\
	T(k)e^{ikx}& \for & x\geq a+\ell,
	\end{array}\right.
	\label{left}\\
	\psi_r(x;k)&=&\left\{\begin{array}{ccc}
	T(k)e^{-ikx} & \for & x\leq a,\\
	R^r(k)e^{ikx}+e^{-ikx} & \for & x\geq a+\ell,
	\end{array}\right.
	\label{right}
	\eea
where $R^{l/r}(k)$ and $T(k)$ are respectively the left/right reflection and transmission amplitudes of the system. They are related to the entries $M_{ij}(k)$ of the transfer matrix via
	\begin{align}
	&R^l(k)=-\frac{M_{21}(k)}{M_{22}(k)},
	&&R^r(k)=\frac{M_{12}(k)}{M_{22}(k)},
	&&T(k)=\frac{1}{M_{22}(k)}.
	\label{RRT}
	\end{align}
The solution of the scattering problem for the permittivity profile (\ref{e1}), equivalently the
potential (\ref{v=}), means the determination of the corresponding reflection and transmission amplitudes, $R^{l/r}(k)$ and $T(k)$, or its transfer matrix $\bM(k)$.

Low-energy potential scattering amounts to the study of the behavior of $R^{l/r}(k)$ and $T(k)$ in the limit $k\to 0$, i.e., for wavenumbers $k$ that are much smaller than the inverse of the largest relevant length scale entering the expression for the potential. For the cases where $v$ does not depend on $k$, there is a well-established mathematical theory describing the $k\to 0$ asymptotic behavior of $R^{l/r}(k)$ and $T(k)$, \cite{bolle-1985,bolle-1987,newton-1986,klaus-1988,aktosun-2001,yafaev}. In particular, for exponentially decaying and finite-range potentials, this theory establishes the analyticity of $R^{l/r}(k)$ and $T(k)$ at $k=0$, and provides iterative schemes for computing the coefficients of their Taylor expansions \cite{bolle-1985,p164}. It is important to note that these results rely on the $k$-independence of the potential. Therefore they do not apply to potentials of the from (\ref{v=}) that are associated with the Helmholtz equation (\ref{H-eq}). 

The purpose of the present article is to introduce an alternative approach to low-frequency scattering that is tailored to deal with the Helmholz equation~(\ref{H-eq}). Similarly to the treatment of \cite{p164}, it relies on a recently developed dynamical formulation of stationary scattering in one dimension in which the transfer matrix is expressed in terms of the time-evolution operator for a non-unitary two-level quantum system \cite{ap-2014,pra-2014a}. Surprisingly, this approach turns out to be more powerful than the ones applicable for the $k$-independent finite-range potentials, because rather than offering an iterative scheme for computing the coefficients of the low-energy series for $R^{l/r}(k)$ and $T(k)$, it gives formulas for the coefficients of the low-energy series for the transfer matrix $\bM(k)$.  

The organization of the article is as follows. In Sec.~\ref{S2} we review a dynamical formulation of the stationary scattering defined by the Helmholtz equation (\ref{H-eq}) that expresses the transfer matrix $\bM(k)$ in terms of the evolution operator for a certain non-unitary two-level quantum system. In Sec.~\ref{S3} we present a mathematical identity which allows us to determine the structure of the terms appearing in the Dyson series expansion of the evolution operator for this system. This in turns paves the way toward identifying the low-frequency series expansion of $\bM(k)$ and determining its coefficients. In Sec.~\ref{S4} we examine the low-frequency behavior of the reflection and transmission amplitudes and discuss some of the physical implications of our findings. 
In Sec.~\ref{S5}, we extend the domain of the application of our method to the scattering in the half-line, $[0,+\infty)$. Here we consider a general scenario where the scattering problem is determined by the permittivity profile of the slab and the choice of a homogeneous boundary condition at $x=0$. Finally in Sec.~\ref{S6}, we summarize our main results and present our concluding remarks.

\section{Dynamical formulation of stationary scattering in 1D}
\label{S2}

Consider the two-component state vector \cite{pra-2014a,jpa-2014a},
	\begin{align}
	&\bPsi(x;k):=\frac{1}{2}\left[\begin{array}{c}
	\psi(x;k)-ik^{-1}\partial_x\psi(x;k)\\
	\psi(x;k)+ik^{-1}\partial_x\psi(x;k)\end{array}\right].
	\label{Psi=}
	\end{align}
It is easy to check that $\psi(k;x)$ is a solution of the Helmholtz equation~(\ref{H-eq}) if and only if $\bPsi(x;k)$ satisfies the time-dependent Schr\"odinger equation,
	\be
	i\partial_x\bPsi(x;k)=\bcH(x;k)\bPsi(x;k),
	\label{sch-eq-TD}
	\ee
where $x$ plays the role of time, 	
	\begin{align}
	&\bcH(x;k):=
	-\frac{k}{2}\Big[\hat\varepsilon(x;k)\,\bcK+\bcK^T\Big],
	\label{H=}
	\end{align}
	\be
	\bcK:=\left[\begin{array}{cc}
	1 & 1\\
	-1 & -1\end{array}\right]=i\bsigma_2+\bsigma_3,
	\label{bcK}
	\ee
$\bcK^T$ labels the transpose of $\bcK$, and $\bsigma_j$ are the Pauli matrices;
	\begin{align}
	&\bsigma_1:=\left[\begin{array}{cc}
	0 & 1\\
	1 &0\end{array}\right],
	&&\bsigma_2:=\left[\begin{array}{cc}
	0 & -i\\
	i &0\end{array}\right],
	&&\bsigma_3:=\left[\begin{array}{cc}
	1 & 0\\
	0 & -1\end{array}\right].
	\label{Pauli}
	\end{align}
	
Let $\bcU(x,x_0;k)$ denote the evolution operator for the Hamiltonian~$\bH(x;k)$, where $x_0$ is an initial `time.' By definition, it is the unique solution of the initial-value problem, 
	\begin{align}
	&i\partial_x\,\bcU(x,x_0;k)=\bcH(x;k)\,\bcU(x,x_0;k),
	&&\bcU(x_0,x_0;k)=\bI,
	\label{sch-eq-U}
	\end{align}
where $\bI$ labels the $2\times 2$ identity matrix. It is often useful to turn (\ref{sch-eq-U}) into an integral equation and use it to express its formal solution as the
following Dyson series \cite{weinberg}.
    \bea
    \bcU(x,x_0;k)&:=&\bI+\sum_{n=1}^\infty(-i)^n\!\!\int_{x_0}^x\!\!dx_n
    \int_{x_0}^{x_n}\!\!\!\! dx_{n-1}\cdots\int_{x_0}^{x_2}\!\!\!\!dx_1
    \bcH(x_n;k)\bcH(x_{n-1};k)\cdots\bcH(x_1;k).~~~~
    \label{U-def}
    \eea
In view of (\ref{sch-eq-TD}) and (\ref{sch-eq-U}), $\bcU(x,x_0;k)$ evolves the state vector $\bPsi(x_0;k)$ according to
	\be
	\bcU(x,x_0;k)\bPsi(x_0;k)=\bPsi(x;k).
	\label{evolve}
	\ee
Because $\bcH(x;k)$ is a non-Hermitian matrix, $\bcU(x,x_0;k)$ fails to be unitary, and (\ref{evolve}) describes the dynamics of a non-unitary two-level quantum system.

Let us introduce $a_-:=a$ and $a_+:=a+\ell$. Then, we can use (\ref{asym}) and (\ref{Psi=}) to show that
	\be
	\bPsi(a_\pm;k)=\left[\begin{array}{c}
	A_\pm(k) e^{ika_\pm}\\
	B_\pm(k) e^{-ika_\pm}\end{array}\right]=
	e^{ika_\pm\bsigma_3}\left[\begin{array}{c}
	A_\pm(k)\\
	B_\pm(k)\end{array}\right].
	\label{Psi=a} 
	\ee
Setting $x_0=a_-$ and $x=a_+$ in (\ref{evolve}) and using (\ref{Psi=a}) in the resulting equation, we find
	\[\left[\begin{array}{c}
	A_+(k)\\
	B_+(k)\end{array}\right]=
	e^{-ika_+\bsigma_3}\bcU(a_+,a_-;k)\,e^{ika_-\bsigma_3}
	\left[\begin{array}{c}
	A_-(k)\\
	B_-(k)\end{array}\right].\]
Comparing this equation with (\ref{M-def}) and recalling that $\bM(k)$ is the unique 
$A_-(k)$- and $B_-(k)$-independent matrix satisfying (\ref{M-def}), we conclude that
	\be
	\bM(k)=e^{-ika_+\bsigma_3}\bcU(a_+,a_-;k)e^{ika_-\bsigma_3}.
	\label{M=1}
	\ee
In the absence of the slab, $\hat\varepsilon(x;k)=1$, $\bcH(x;k)=-k\bsigma_3$, $\bcU(a_+,a_-;k)=\exp(ik\ell\bsigma_3)$, and (\ref{M=}) gives $\bM(k)=\bI$.
	
Next, consider the translation $x\to\tilde x:=x-a$, and set $\tilde v(\tilde x;k):=v(x;k)$, so that $\tilde v(x)=v(x+a;k)$. Then according to (\ref{v=}), $\tilde v(x)=0$ for $x\notin[0,\ell]$. Therefore it describes an identical copy of our slab that is placed between the planes $x=0$ and $x=\ell$. In view of (\ref{asym}) and (\ref{M-def}), under the translation $x\to x-a$, the transfer matrix transforms according to \cite{bookchapter},
	\begin{align}
	&\bM(k)\to \tilde\bM(k)=e^{-ika\bsigma_3}\bM(k)e^{ika\bsigma_3},
	\label{M-trans}
	\end{align}
where $\tilde\bM(k)$ is the transfer matrix of the potential $\tilde v(x;k)$ that is supported in $[0,\ell]$. Using the inverse of the transformation rule (\ref{M-trans}), we can determine the transfer matrix of our slab to that of its translated copy residing in $[0,\ell]$. 

In the following, first we set $a_-=a=0$ and $a_+=\ell$ which turns (\ref{M=1}) into
	\be
	\bM(k)=e^{-ik\ell\bsigma_3}\bcU(\ell,0;k).
	\label{M=}
	\ee
We then explore the low-frequency properties of this transfer matrix and obtain those of our original slab, with $a\neq 0$, by performing the inverse of the transformations (\ref{M-trans}). In terms of the entries of the transfer matrix, this takes the form,
	\begin{align}
	&M_{11}(k)\to M_{11}(k),
	&&M_{12}(k)\to e^{2ika} M_{12}(k),
	\label{M1-translate}\\
	&M_{21}(k)\to e^{-2ika} M_{21}(k),
	&&M_{22}(k)\to \tilde M_{22}(k).
	\label{M2-translate}
	\end{align}
In view of (\ref{RRT}), these correspond to
	\begin{align}
	& R^{l}(k)\to e^{-2ika} R^{l}(k),
	&& R^{r}(k)\to  e^{2ika} R^{r}(k),
	&& T(k)\to T(k).
	\label{RRT-trans}
	\end{align}
	
\section{Low-frequency series for the transfer matrix}
\label{S3}

Because the Hamiltonian matrix (\ref{H=}) is proportional to $k$, the terms of 
the Dyson series (\ref{U-def}) are proportional to nonnegative integer powers of $k$. This observation together with Eq.~(\ref{M=}) provide a pathway towards constructing a low-frequency series expansion for the transfer matrix. The following mathematical result provides the key for determining the coefficients of this series.
\begin{itemize}
\item[]{\bf Lemma}: Let $\{\fa_j\}$ and $\{\fb_j\}$ be sequences of complex numbers whose terms are labeled by positive integers $j$,  $\bcH_j:=\frac{1}{2}\big(\fa_j\bcK+\fb_j\bcK^T\big)$, and $\bsigma_\pm:=\bI\pm\bsigma_1$. Then for every positive integer $n$,
	\bea
	\bcH_{2n}\bcH_{2n-1}\cdots\bcH_1&=&
	\frac{1}{2}\left(\prod_{j=1}^n \fa_{2j} \fb_{2j-1}~\bsigma_-+
	\prod_{j=1}^n \fb_{2j} \fa_{2j-1}~\bsigma_+\right),
	\label{even}\\
	\bcH_{2n+1}\bcH_{2n}\cdots\bcH_1&=&
	\frac{1}{2}\left(\fa_{2n+1}\prod_{j=1}^n \fb_{2j} \fa_{2j-1}~\bcK+
	\fb_{2n+1}\prod_{j=1}^n \fa_{2j} \fb_{2j-1}~\bcK^T\right).
	\label{odd}
	\eea
\item[]{\bf Proof}: In view of (\ref{bcK}) and (\ref{Pauli}), $\bcK^2=\bcK\bsigma_-=\bcK^T\bsigma_+=\bzero$, $\bcK\bcK^T=2\bsigma_-$, $\bcK^T\bcK=2\bsigma_+$,
	$\bcK\bsigma_+=2\bcK$,  and $\bcK^T\bsigma_-=2\bcK^T$. Eqs.~(\ref{even}) and (\ref{odd}) follow from these relations by induction on $n$.
	\end{itemize}
	
Now, let $x_j\in\R$ and set $\fa_j:=-k\,\hat\varepsilon(x_j;k)$ and $\fb_j:=-k$. Then $\bcH_j:=\bcH(x_j;k)$, and (\ref{even}) and (\ref{odd}) give 
	\bea
	\bcH(x_{2n};k)\bcH(x_{2n-1};k)\cdots\bcH(x_1;k)&=&
	\frac{k^{2n}}{2}
	\left(\prod_{j=1}^n \hat\varepsilon(x_{2j};k)~\bsigma_-+
	\prod_{j=1}^n \hat\varepsilon(x_{2j-1};k)~\bsigma_+\right),\\
	\bcH(x_{2n+1};k)\bcH(x_{2n};k)\cdots\bcH(x_1;k)&=&
	-\frac{k^{2n+1}}{2}
	\left(\prod_{j=1}^{n+1} \hat\varepsilon(x_{2j-1};k)~\bcK+
	\prod_{j=1}^n \hat\varepsilon(x_{2j};k)~\bcK^T\right).~~~
	\eea
Substituting these equations in (\ref{U-def}), we find 
	\bea
	\bcU(x,x_0;k)&=&\bI+\frac{1}{2}\sum_{n=1}^\infty(-1)^n
	\Big\{-ik^{2n-1}\big[p_{2n-1}(x,x_0;k)\,\bcK+q_{2n-1}(x,x_0;k)\,\bcK^T\big]+
	\nn\\
	&&\hspace{3.2cm}
	k^{2n}\big[ p_{2n}(x,x_0;k)\,\bsigma_-+q_{2n}(x,x_0;k)\,\bsigma_+\big]
	\Big\},
	\label{U=}
	\eea
where
	\begin{align}
	&p_1(x,x_0;k):=\int_{x_0}^xdx_1\,\hat\varepsilon(x_1;k),	
	&&q_1(x,x_0;k):=x-x_0,
	\label{pq-1}
	\end{align}
for $n\geq 1$,
{\small	\bea
	p_{2n+1}(x,x_0;k)&:=&\int_{x_0}^x\!\!dx_{2n+1}
    	\int_{x_0}^{x_{2n+1}}\!\!\!\!\! dx_{2n}\cdots
	\int_{x_0}^{x_3}\!\!\!\!dx_2
	\int_{x_0}^{x_2}\!\!\!\!dx_1
	\prod_{j=1}^{n+1} \hat\varepsilon(x_{2j-1};k)
	\label{p2n+1=}\\
	&=&\int_{x_0}^x\!\!\!dx_{2n+1}
    	\int_{x_0}^{x_{2n-1}}\!\!\!\! dx_{2n-1}\cdots
	\int_{x_0}^{x_5}\!\!\!\!dx_3
	\int_{x_0}^{x_3}\!\!\!\!dx_1
	\hat\varepsilon(x_{2n+1};k)
	\prod_{j=1}^{n}(x_{2j+1}-x_{2j-1})\hat\varepsilon(x_{2j-1};k),\nn\\
	q_{2n+1}(x,x_0;k)&:=&
	\int_{x_0}^x\!\!dx_{2n+1}
    	\int_{x_0}^{x_{2n+1}}\!\!\!\!\! dx_{2n}\cdots
	\int_{x_0}^{x_3}\!\!\!\!dx_2
	\int_{x_0}^{x_2}\!\!\!\!dx_1
	\prod_{j=1}^n \hat\varepsilon(x_{2j};k)
	\label{q2n+1=}\\
	&=&	\int_{x_0}^x\!\!\!\!dx_{2n}
    	\int_{x_0}^{x_{2n}}\!\!\!\! dx_{2n-2}\cdots
	\int_{x_0}^{x_6}\!\!\!\!dx_4
	\int_{x_0}^{x_4}\!\!\!\!dx_2(x-x_{2n})
	\prod_{j=1}^n(x_{2j}-x_{2j-2})\hat\varepsilon(x_{2j};k),\nn
	\eea
	\bea
	p_{2n}(x,x_0;k)&:=&\int_{x_0}^x\!\!dx_{2n}
    	\int_{x_0}^{x_{2n}}\!\!\!\!\! dx_{2n-1}\cdots
	\int_{x_0}^{x_3}\!\!\!\!dx_2
	\int_{x_0}^{x_2}\!\!\!\!dx_1
	\prod_{j=1}^n \hat\varepsilon(x_{2j};k)
	\label{Pn=}\\
	&=&	\int_{x_0}^x\!\!dx_{2n}
    	\int_{x_0}^{x_{2n}}\!\!\!\! dx_{2n-2}\cdots
	\int_{x_0}^{x_6}\!\!\!\!dx_4
	\int_{x_0}^{x_4}\!\!\!\!dx_2
	\prod_{j=1}^n(x_{2j}-x_{2j-2})\hat\varepsilon(x_{2j};k),\nn\\
	q_{2n}(x,x_0;k)&:=&\int_{x_0}^x\!\!dx_{2n}
    	\int_{x_0}^{x_{2n}}\!\!\!\!\! dx_{2n-1}\cdots
	\int_{x_0}^{x_3}\!\!\!\!dx_2
	\int_{x_0}^{x_2}\!\!\!\!dx_1
	\prod_{j=1}^n \hat\varepsilon(x_{2j-1};k)
	\label{Qn=}\\
	&=&	\int_{x_0}^x\!\!\!dx_{2n-1}
    	\int_{x_0}^{x_{2n-1}}\!\!\!\! dx_{2n-3}\cdots
	\int_{x_0}^{x_5}\!\!\!\!dx_3
	\int_{x_0}^{x_3}\!\!\!\!dx_1
	\prod_{j=1}^{n}(x_{2j+1}-x_{2j-1})\hat\varepsilon(x_{2j-1};k)\Big],\nn
	\eea}
and in (\ref{Qn=}), $x_{2n+1}:=x$.
	
Next, consider the case where $x=\ell$ and $x_0=0$. Then the variables $x_j$ of the integrals appearing in (\ref{pq-1}) -- (\ref{Qn=}) range over the interval $[0,\ell]$, and (\ref{e1}) with $a=0$ gives	
	\be
	\hat\varepsilon(x_j;k)=w(\hat x_j;k),
	\label{e1-w}
	\ee
where $\hat x_j:=x_j/\ell$. If we make the change of variables of integration, $x_j\to\hat x_j$ in (\ref{pq-1}) -- (\ref{Qn=}) and substitute (\ref{e1-w}) in the result, we find that $p_n(\ell,0;k)$ and $q_n(\ell,0;k)$ are proportional to $\ell^n$. In particular, introducing 
	\begin{align}
	&r_0^+(k):=1, \quad\quad\quad r_0^-(k):=0, \quad\quad\quad 
	r_{n}^\pm(k):=\frac{p_n(\ell,0;k)\pm q_n(\ell,0;k)}{2\ell^n}~~\for~~n>1,
	\label{rn-def}
	\end{align}
and making use of (\ref{M=}) and (\ref{U=}), we find
	\bea
	\bM(k)&=&e^{-ik\ell\bsigma_3}\Big[\bI+\sum_{n=1}^\infty 
	(ik\ell\,\bsigma_3)^n\bS_n(k)\Big]
	\label{M=2a}\\
	&=&\bI+e^{-ik\ell\bsigma_3}\sum_{n=1}^\infty
	(ik\ell\,\bsigma_3)^n\Big[\bS_n(k)-\frac{1}{n!}\,\bI\Big],\\
	&=&\bI+\sum_{n=1}^\infty (-ik\ell)^n\:\bm_n(k),
	\label{M=2}
	\eea
where 
	\begin{align}
	&\bS_{n}(k):=r_{n}^+(k)\,\bI-(-1)^nr^-_{n}(k)\,\bsigma_1,
	\label{S=}
	\\[6pt]
	&\bm_{n}(k):=
	\bsigma_3^n\big[t^+_{n}(k)\bI-(-1)^nt^-_{n}(k)\bsigma_1\big],
	\label{mn-def}\\
	&t_n^\pm(k):=\sum_{m=0}^n\frac{(\mp 1)^m\,r_m^\pm(k)}{(n-m)!}.
	\label{tnpm}
	\end{align}
In view of (\ref{M=2}) and (\ref{mn-def}), the entries of the transfer matrix are given by	\begin{align}
	&M_{11}(k)=1+\sum_{n=1}^\infty t^+_n(k)\,(-ik\ell)^n,
	&&M_{12}(k)=-\sum_{n=1}^\infty t^-_n(k)\,(-ik\ell)^n,
	\label{M11-M12}\\
	&M_{21}(k)=-\sum_{n=1}^\infty t^-_n(k)\,(ik\ell)^n,
	&&M_{22}(k)=1+\sum_{n=1}^\infty  t^+_n(k)\,(ik\ell)^n.
	\label{M21-M22}
	\end{align}
	
Eq.~(\ref{M=2}), (\ref{M11-M12}), and (\ref{M21-M22}) give the low-frequency series expansions for the transfer matrix and its entries provided that we identify the term ``low frequency'' with the condition ``$k\ell\ll 1$.'' 	This is a reasonable choice, for $\ell$ is the natural length scale for the problem. Notice also that for situations where we can ignore the effects of dispersion, $t_n^\pm$ become $k$-independent, and 
(\ref{M11-M12}) and (\ref{M21-M22}) yield Taylor series in powers of $k$. In general, by low-frequency expansion we mean a series expansion in integer powers of $k\ell$. The coefficients of this series can depend on $k$, for $k$ and $k\ell$ as independent variables. 

Next, we examine the convergence properties of the low-energy series expansions for $M_{ij}(k)$. Because $e^{\pm ik\ell}$ is an entire function, we can confine our attention to the entries of the series in the square bracket on the right-hand side of (\ref{M=2a}). According to (\ref{S=}) the $n$-th term of these series are bounded by $|r_n^\pm(k)|$. To determine the large-$n$ behavior of the latter, we use (\ref{bound-zero}) and (\ref{p2n+1=}) -- (\ref{e1-w}) to show that
	\begin{align}
	&|p_{2n}(\ell,0;k)|\leq \frac{(b\ell)^{2n}}{(2n)!},
	&&|q_{2n}(\ell,0;k)|\leq \frac{(b\ell)^{2n}}{(2n)!},\nn\\
	&|p_{2n+1}(\ell,0;k)|\leq \frac{b(b\ell)^{2n+1}}{(2n+1)!},
	&&|q_{2n+1}(\ell,0;k)|\leq \frac{(b\ell)^{2n+1}}{(2n+1)!\, b}.\nn
	\end{align}
These together with (\ref{rn-def}) and the fact that $n!\geq\sqrt{2\pi n} (n/e)^{n}$ imply
	\be
	|r_n^\pm(k)|\leq\frac{|p_n(\ell,0;k)|+|q_n(\ell,0;k)|}{2\ell^n}<
	\frac{(b^2+1)b^{n}}{2b\:n!}\leq
	\frac{b^2+1}{2b\sqrt{2\pi n}}\left(\frac{e\, b}{n}\right)^{\!\!n}.
	\label{bound-r}
	\ee
This relation shows that the series on the right-hand side of (\ref{M=2a}) and consequently the low-frequency series for the entries of the transfer matrix, i.e.,  the right-hand sides of
(\ref{M11-M12}) and (\ref{M21-M22}), have an infinite radius of convergence. Furthermore, the truncation of these series that involves neglecting terms of order $n$ and higher in powers of $k\ell$ leads to reliable approximate expressions for $M_{ij}(k)$ provided that
	\be
	k\ell\ll  \left(\frac{2b\sqrt{2\pi n}}{b^2+1}\right)^{\!\!1/n}\!\!\frac{n}{e\, b}~~~\mbox{or}~~~
	\frac{\ell}{\lambda}\ll
	\left(\frac{2b\sqrt{2\pi n}}{b^2+1}\right)^{\!\!1/n}\!\!\frac{n}{2\pi e\, b},
	\label{bound-3}
	\ee
where $\lambda:=2\pi/k$ is the wavelength. According to (\ref{bound-3}), the knowledge of a few of the terms on the right-hand sides of (\ref{M11-M12}) and (\ref{M21-M22}) provides an accurate description of the scattering properties of sufficiently thin inhomogeneous films, where $\ell\ll\lambda$, and slabs made out of the epsilon-near-zero metamaterials \cite{alu-2007,Ni-2020}, where $b\ll 1$.

We can extend the above analysis to situations where the relative permittivity is unbounded but piecewise continuous, i.e., $w(\hat x,k)$ are piecewise continuous functions of $\hat x\in[\frac{a}{\ell},\frac{a}{\ell}+1]$ and $k\in\R$. In this case for each $\fK\in\R^+$, there is a positive number $b_\fK$ such that $|w(\hat x,k)|\leq b_\fK$ for all $\hat x\in[\frac{a}{\ell},\frac{a}{\ell}+1]$ and $k\in[-\fK,\fK]$. Under this condition, (\ref{bound-r}) holds for $b=b_\fK$ and $k\in[-\fK,\fK]$, and the radius of convergence of the low-frequency series for the entries of the transfer matrix is not smaller than $\fK\ell$.

\section{Reflection and transmission of low-frequency waves}
\label{S4}

To determine the low-frequency series for the reflection and transmission amplitudes, we need to substitute (\ref{M11-M12}) and (\ref{M21-M22}) in (\ref{RRT}), invert the series for $M_{22}(k)$, and multiply the result by the series for $M_{12}(k)$ and $M_{22}(k)$. For practical applications, this task reduces to the division of certain polynomials in $k\ell$. 

For $n\leq 2$, the calculation of $p_n(\ell,0;k)$, $q_n(\ell,0;k)$, $r^\pm_n(k)$, and $t^\pm_n(k)$ is very easy. Here we report the values of the latter:
	\begin{align}
	&t_1^\pm(k)=\mp\frac{u_0(k)}{2}, 
	&&t^+_2(k)=0, && t^-_2(k)=u_1(k),
	\label{t12}
	\end{align}
where 
	\be
	u_m(k):=\int_0^1d\hat x\,\hat x^m[w(\hat x)-1]=
	\frac{1}{\ell^{m+1}}\int_0^\ell dx\, x^m[\hat\varepsilon(x)-1].
	\label{um}
	\ee
Substituting (\ref{t12}) in (\ref{M21-M22}), we have
	\begin{align}
	&M_{11}(k)=1+\frac{i u_0(k)\,k\ell}{2}+O(k\ell)^3,
	&&~~~~M_{12}(k)=\frac{i u_0(k)\,k\ell}{2}+u_1(k)(k\ell)^2+O(k\ell)^3,
	\label{M1-expand-3}\\
	&M_{21}(k)=-\frac{i u_0(k)\,k\ell}{2}+u_1(k)(k\ell)^2+O(k\ell)^3,
	&&~~~~M_{22}(k)=1-\frac{i u_0(k)\,k\ell}{2}+O(k\ell)^3.
	\label{M2-expand-3}
	\end{align}
These together with (\ref{RRT}) imply
	\bea
	R^l(k)&=&\frac{i u_0(k)\,k\ell}{2}-\left[u_1(k)+\frac{u_0(k)^2}{4}\right](k\ell)^2+O(k\ell)^3,
	\label{RL=2}\\
	R^r(k)&=&\frac{i u_0(k)\,k\ell}{2}+\left[u_1(k)-\frac{u_0(k)^2}{4}\right](k\ell)^2+O(k\ell)^3,
	\label{RR=2}\\
	T(k)&=&1+\frac{i u_0(k)\,k\ell}{2}-\frac{u_0(k)^2 (k\ell)^2}{4}+O(k\ell)^3.
	\label{T=2}
	\eea	
	
The following are some of the physical consequences of (\ref{RL=2}) -- (\ref{T=2}).
	\begin{enumerate}
	\item Knowledge of $u_0(k)$ and $u_1(k)$ is sufficient to determine the low-frequency expression for the reflection coefficients $|R^{l/r}(k)|^2$ up to an including the terms of order $(k\ell)^3$. 
	\item If $(k\ell)^2$ is negligibly small, i.e., (\ref{bound-3}) holds for $n=2$, the reflection non-reciprocity, $|R^{l}(k)|\neq |R^{r}(k)|$, goes away, i.e., it is a quadratic or higher-order effect in powers of $k\ell$.
		\item Suppose that $u_0(k)=0$. Then
		\be
		\frac{1}{\ell}\int_0^\ell dx\,\hat\varepsilon(x;k)=1,
		\label{transparent}
		\ee
and (\ref{T=2}) implies $T(k)=1+O(k\ell)^3$. This means that {\em the slab is transparent for the low frequencies at which its average permittivity equals that of the vacuum.} This observation suggests that coating a sufficiently thin film with a layer of epsilon-near-zero metamaterial can make it transparent. For example, let $\ell_1$ and $\hat\epsilon_1(x;k)$ respectively label the thickness and the relative permittivity of the film, and $b_1$ be the maximum of $|\hat\epsilon_1(x;k)|$. Coating this film by a layer of an epsilon-near-zero metamaterial with thickness $\ell_2:=\ell-\ell_1$ and relative permittivity $\hat\epsilon_2(x;k)$ will make the film transparent provided that we place the film and the coating in the regions of the space corresponding to $0\leq x\leq \ell_1$ and $\ell_1\leq x\leq \ell_1+\ell_2$, and make sure that
	\[\int_{\ell_1}^{\ell_1+\ell_2}dx\,\hat\varepsilon_2(x;k)=
	\ell-\int_{0}^{\ell_1}dx\,\hat\varepsilon_1(x;k),\]
and (\ref{bound-3}) holds for $n=3$ and $b=b_1$. Note that we can achieve the same goal by exchanging the positions of the film and the coating, or coat both faces of the film with epsilon-near-zero layers in such a way that (\ref{transparent}) holds.

	\item At low-frequencies the conditions, $R^{l/r}(k)=O(k\ell)^3$ and $T(k)=1+O(k\ell)^3$, which characterize bidirectional invisibility, are equivalent to $u_0(k)=u_1(k)=0$, i.e., in addition to (\ref{transparent}), we need to impose,
	\be
	\frac{1}{\ell}\int_0^\ell dx\,x\,\hat\varepsilon(x;k)=\frac{\ell}{2}.
	\label{u1-zero}
	\ee

	\item Reflectionlessness at low frequencies from the left or right requires $u_0(k)=u_1(k)=0$. Because this is the same as the condition for bidirectional invisibility, we conclude that {\em unidirectional reflectionlessness and invisibility} \cite{lin,pra-2013a} {\em are forbidden at low-frequencies}. 
	
	\item The condition (\ref{transparent}) which ensures the transparency of the slab and is necessary for its reflectionlessness (and invisibility) implies that the average of the imaginary part of slabs permittivity vanishes;
	\be
	\int_0^\ell dx\,\IM[\hat\varepsilon(x;k)]=0.
	\label{balanced}
	\ee
This shows that, {\em in order for the slab to display transparency, reflectionlessness, or invisibility for low-frequency TE waves, it must have balanced gain and loss, i.e., (\ref{balanced}) holds.} This is the characteristic feature of $\cP\cT$-symmetric material \cite{konotop-review,feng-review}, although there are a variety of non-$\cP\cT$-symmetric material that also contain balanced gain and loss.	
	\end{enumerate}

\section{Low-frequency scattering for Helmholtz equation in the half-line}
\label{S5}

Suppose that the slab we are considering is placed in the half-space $\cS^+$ given by $x\geq 0$, the complementary half-space $\cS^-$ corresponding to $x< 0$ is filled by certain material with planar symmetry, and we are interested in the scattering of right-incident TE waves, i.e., those whose source is placed at $x=+\infty$. The simplest example is a slab that is placed at some distance $a$ from a perfect mirror. 

We can model the above setup in terms of a scattering problem defined on the half-line, $\R^+\cup\{0\}$, by the Helmholtz equation,
	\be
	\partial_x^2\psi(x;k)+k^2\hat\varepsilon(x;k)\psi(x;k)=0,~~~~~x>0,
	\label{H-eq-half}
	\ee
and a boundary condition at $x=0$ that encodes the information about the response of the medium filling $\cS^-$. Specifically, we consider homogenous boundary conditions of the Robin type,
	\be
	\alpha(k)\psi(0;k)+k^{-1}\beta(k)\partial_x\psi(0;k)=0,
	\label{BC}
	\ee
where $\alpha(k)$ and $\beta(k)$ are real- or complex-valued functions satisfying $|\alpha(k)|+|\beta(k)|\neq 0$. Eq.~(\ref{BC}) yields the Dirichlet and Neumann boundary conditions for $\beta(k)=0$ and $\alpha(k)=0$, respectively. The scattering solutions of (\ref{H-eq-half}) and (\ref{BC}) fulfill the asymptotic boundary condition,
	\be
	\psi(x;k)\to e^{-ikx}+ \cR(k) e^{ikx}~~~\for~~~x\to+\infty,
	\nn
	\ee
where $e^{-ikx}$ represents the incident wave, and $\cR(k)$ is the reflection amplitude. 

Ref.~\cite{ap-2019b} offers a simple method of extending the utility of the transfer matrix of potential scattering in the full line to the scattering problems in the half-line. Specifically, it relates the reflection amplitude $\cR(k)$ of the latter to the entries of the transfer matrix for its trivial extension to the full line. Employing this scheme for dealing with our one-dimensional electromagnetic scattering problem in the half-line yields,
	\be
	\cR(k)=\frac{M_{11}(k)-\gamma(k)M_{12}(k)}{M_{21}(k)-\gamma(k)M_{22}(k)},
	\label{cR=}
	\ee
where $M_{ij}(k)$ are the entries of the transfer matrix for the same system in the absence of the material filling $\cS^-$, and
	\[\gamma(k):=\frac{\alpha(k)+i\beta(k)}{\alpha(k)-i\beta(k)}.\]
The choice, $\alpha(k)=1$ and $\beta(k)=-i$, corresponds to the situation where the half-space $\cS^-$ is empty. In this case, $\gamma(k)=\infty$, the right-hand side of (\ref{cR=}) coincides with the right reflection amplitude of the slab, $R^r(k)$, and (\ref{cR=expand}) reproduces the low-frequency expansion for the latter, i.e., (\ref{RR=2}) with an extra phase factor $e^{2ika}$ multiplying its right-hand side,
which reflects the fact that $a\neq 0$. 	
	
Eq.~(\ref{cR=}) reduces the study of the low-frequency behavior of the reflection amplitude $\cR(k)$ to that of the transfer matrix of the slab system we considered in Sec.~\ref{S3} except that now the slab occupies the region defined by $a\leq x\leq a+\ell$ for some $a\geq 0$. According to (\ref{M1-translate}) and (\ref{M2-translate}), $M_{11}(k)$ and $M_{22}(k)$ do not depend on $a$, while the expressions we have obtained for $M_{12}(k)$ and $M_{21}(x)$ in Sec.~\ref{S3} acquire extra phase factors $e^{2ika}$ and $e^{-2iak}$, respectively. Furthermore, it is not difficult to see that the quantities $u_m(k)$ associate with a slab with $a>0$ are given by
	\[ u_m(x)=\frac{1}{\ell^{m+1}}\int_{a}^{a+\ell}\!\!\! dx\:(x-a)^m[\hat\varepsilon(x;k)-1].\]
Taking these into account and using 
(\ref{M11-M12}) and (\ref{M21-M22}) we can obtain the low-frequency expansion of $\cR(k)$ as the ratio of a pair of power series in $k\ell$. In particular, by virtue of (\ref{M1-expand-3}) and (\ref{M2-expand-3}), we have
	\bea
	\cR(k)&=&-\frac{1}{\gamma(k)}
	+\frac{i}{2}\xi_1(k)^2 u_0(k)\,k\ell+
	\left[\xi_2(k)u_1(k)-\frac{1}{4}e^{-iak}\xi_1(k)^3u^2_0(k)\right](k\ell)^2
	+O(k\ell)^3.~~~
	\label{cR=expand}
	\eea
where
	\[\xi_m(k):=e^{imak}-e^{-imak}\gamma(k)^{-m}.\]
For the Dirichlet and Neumann boundary conditions, which respectively correspond to $\gamma(k)=1$ and $\gamma(k)=-1$, Eq.~(\ref{cR=expand}) simplifies considerably. 

Imposing the Dirichlet boundary condition at $x=0$, which corresponds to placing a perfect mirror ar $x=0$, we find
	\bea
	\cR(k)&=&-1-2i\sin^2(ak)u_0(k)\,k\ell+2i
	\Big[\sin(2ak)u_1(k)+\nn\\
	&&\hspace{5.45cm}e^{-iak}\sin^3(ak)u_0(k)^2\Big](k\ell)^2+O(k\ell)^3.
	\label{cR=Dirichet}
	\eea
The following are consequences of this equation.
	\begin{enumerate}
	\item If $ak$ is an integer multiple of $\pi$, $\cR(k)=-1+O(k\ell)^3$. This means that {\em if we adjust the distance $a$ between the slab and the mirror to be an integer multiple of the half-wavelength, the slab becomes reflectionless at low frequencies irrespectively of the details of its permittivity profile.} 
	
	\item The absorption coefficient, $\cA(k):=1-|\cR(k)|^2$, admits the following low-energy expansion.
	\bea
	\cA(k)&=&4\,\IM[u_0(k)]\sin^2(ak)\,k\ell-8\sin(ak)
	\Big\{\RE[u_0(k)]\,\IM(u_0(k)]\sin^2(ak)\cos(ak)+\nn\\
	&& 
	\IM[u_0(k)]^2\sin^3(ak)+\IM[u_1(k)]\cos(ak)\Big\}(k\ell)^2+O(k\ell)^3,
	\label{ab-co}
	\eea
where ``$\IM$'' and ``$\RE$'' respectively stand for the imaginary and real parts of their arguments. 
If $\IM[u_0(k)]=0$, (\ref{ab-co}) becomes,
	\be
	\cA(k)=-4\,\IM[u_1(k)]\sin(2ak)(k\ell)^2+O(k\ell)^3.
	\label{ab-co-balanced}
	\ee
Therefore, whenever $\IM[u_1(k)]$ takes a nonzero value so that we can neglect the cubic and higher order terms on the right-hand side of (\ref{ab-co-balanced}), the absorption coefficient can take both negative and positive values. This is interesting, because by adjusting the distance $a$ between the slab and the mirror, we can make the slab act as an absorber or amplifier for low-frequency waves. This is more pronounced when $a$ is related to the wavelength $\lambda:=2\pi/k$ according to $a=(2n+1)\lambda/8$, where $n$ is a positive integer.

Notice that because,
	\begin{align}
	&\IM[u_0(k)]=\frac{1}{\ell}\int_a^{a+\ell}\!\!\! dx\,\IM[\hat\varepsilon(x;k)],
	&&\IM[u_1(k)]=\frac{1}{\ell^2}\int_a^{a+\ell}\!\!\! dx\,(x-a)\,\IM[\hat\varepsilon(x;k)],
	\label{u0-u1=}
	\end{align}
$\IM[u_0(k)]$ signifies the average of the imaginary part of the slab's permittivity, and $\IM[u_0(k)]=0$ implies that the slab has balanced gain and loss properties. It is also easy to see that this condition does not conflict $\IM[u_1(k)]\neq 0$. Therefore according to (\ref{ab-co-balanced}),
{\em depending on its position, a slab with balanced loss and gain can serve both as an absorber and amplifier for low-frequency waves.} Typical examples are $\cP\cT$-symmetric slabs placed in a waveguide \cite{muga-2005,prl-2009} or empty space \cite{ge-2011,chong-2011,jpa-2012}. In the remainder of this section we examine an experimentally less stringent system that would allow for a realization of this effect.

\end{enumerate}

\begin{figure}
  	\begin{center}
 	\includegraphics[scale=0.3]{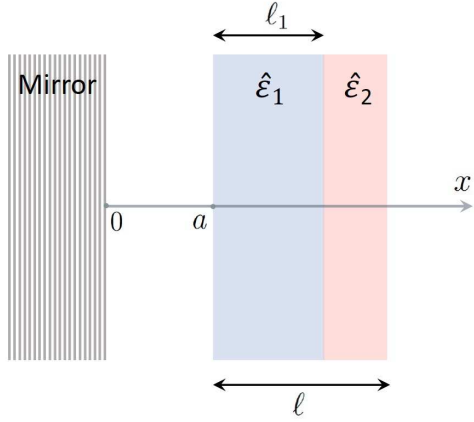}
  	\end{center}
  	\vspace{-12pt}
  	\caption{Schematic view of a bilayer slab with permittivity profile (\ref{bilayer}) placed next to a perfect mirror. }
	\label{fig1}
	\end{figure}

Fig.~\ref{fig1} 
shows a bilayer slab $\sS$ that is placed in front of a perfect mirror and has a relative permittivity profile of the form:
	\be
	\hat\varepsilon(x;k)=\left\{\begin{array}{ccc}
	1 &\for& x<a~~{\rm and}~~x>a+\ell,\\
	\hat\varepsilon_1(k) &\for& a\leq x<a+\ell_1,\\
	\hat\varepsilon_2(k) &\for&a+\ell_1\leq x\leq a+\ell,\end{array}\right.
	\label{bilayer}
	\ee
where $\hat\varepsilon_1$ and $\hat\varepsilon_2$ are the relative permittivities of the layers, and
$\ell_1$ and $\ell_2:=\ell-\ell_1$ are their thickness. Suppose that for some $k$ satisfying $k\ell\ll1$, $\IM[\hat\varepsilon_1(k)]>0$ and $\IM[\hat\varepsilon_2(k)]<0$, so that the closest layer to the mirror is lossy and the other layer is made of a gain material. Introducing the real parameter,
	\be
	\mu:=-\frac{\IM[\hat\varepsilon_2(k)]}{\IM[\hat\varepsilon_1(k)]},\nn
	\ee
and making use of (\ref{u0-u1=}) and (\ref{bilayer}), we can easily show that
	\bea
	\IM[u_0(k)]&=&\IM[\hat\varepsilon_1(k)]\left[\frac{(1+\mu)\ell_1}{\ell}-\mu\right],
	\label{bilayer-u0}\\
	\IM[u_1(k)]&=&\frac{\IM[\hat\varepsilon_1(k)]}{2}\left[
	\frac{(1+\mu)\ell_1^2}{\ell^2}-\mu\right].
	\label{bilayer-u1}
	\eea
According to (\ref{bilayer-u0}), $\sS$ has balanced gain and loss, i.e., $\IM[u_0(k)]=0$, provided that 
$\mu>0$ and the thickness of the layers are given by
	\begin{align}
	&\ell_1=\frac{\ell}{1+\mu^{-1}}, &&\ell_2=\frac{\ell}{1+\mu}.
	\nn
	\end{align}
Substituting the first of these equations in (\ref{bilayer-u1}) and using the result in (\ref{ab-co-balanced}), we obtain
	\be
	\cA(k)=\cA_\star\:\sin(2ka)\,(k\ell)^2+O(k\ell)^3,
	\label{slab-A}
	\ee
where
	\[\cA_\star:=\frac{2\,\IM[\hat\varepsilon_1(k)]\:\mu}{\mu+1}=-\frac{2\,\IM[\hat\varepsilon_2(k)]}{\mu+1}.\]
	
The bilayer slab $\sS$ is $\cP\cT$-symmetric if and only if  $\ell_1=\ell_2=\ell/2$ and $\hat\varepsilon_2(k)=\hat\varepsilon_1(k)^*$. In this case, $\mu=1$ and $\cA_\star=\IM[\hat\varepsilon_1(k)]$. Notice that $\cP\cT$-symmetry also requires $\RE[\hat\varepsilon_2(k)]=\RE[\hat\varepsilon_1(k)]$. Therefore, compared with its $\cP\cT$-symmetric special case, $\sS$ involves two more free parameters, namely $\mu$ and $\RE[\hat\varepsilon_2(k)]$. This would considerably ease an experimental confirmation of Eq.~(\ref{slab-A}) pertaining the oscillatory behavior of the low-frequency absorption coefficient of the slab as a function of its distance to the mirror.

\section{Conclusions}
\label{S6}

Mathematicians have provided a detailed but elaborate construction of the low-frequency series expansions of the reflection and transmission amplitude for exponentially decaying (and hence finite-range) potentials in the 1980's \cite{bolle-1985,bolle-1987}. The dynamical formulation of stationary scattering offers a much simpler and easier-to-use approach for constructing these series \cite{p164}. Similarly to earlier treatments of the subject, this approach cannot be applied for energy-dependent potentials such as those appearing in the study of the Helmholtz equation in one dimension. In the present article, we have offered a comprehensive and essentially self-contained treatment of the low-frequency scattering solutions of the Helmholtz equation for a finite-range scatterer residing in the real line or the half-line.

Our approach relies on two basic developments: 1) A dynamical formulation of the scattering problem that identifies the low-frequency series for the transfer matrix with a Dyson series; 2) A simple and elegant mathematical identity that allows us to derive formulas for the coefficients of this series. The importance of this feature of our approach becomes evident once we realize that the coefficients of the low-energy series for the transfer matrix of an energy-independent finite-range potential can only be computed iteratively in terms of certain solutions of the zero-energy Schr\"odinger equation \cite{bolle-1985,bolle-1987,p164}. 

In our treatment of the low-frequency scattering, we have used the size of the scatterer, which is quantified by the thickness $\ell$ of our slab, as the largest relevant length scale for the scattering problem. This has led us to identify ``$k\ell\ll 1$'' as the physical condition describing the ``smallness of the frequency.'' To assess the reliability of the approximate descriptions of the low-frequency scattering that are obtained by truncating the low-frequency series expansion of the transfer matrix (or reflection and transmission amplitudes), we have obtained error bounds on the neglected terms. These turn out to depend on the maximum value of the modulus of the relative permittivity, i.e., $|\hat\varepsilon|$, and suggest the scattering of TE waves by epsilon-near-zero metamaterial slabs as an area in which we can safely employ these approximations.
 
We have used the low-frequency series for the transfer matrix to obtain those for the reflection and transmission amplitudes. This reveals a number of interesting general predictions on the transparency, reflectionlessness, and invisibility at low frequencies. The same applies for our results on the low-frequency behavior of scattering defined by the Helmholtz equation in the half-line. In particular, we show that a slab displaying balanced gain and loss properties can serve both as an absorber and amplifier for low-frequency waves, if we place it in front of a perfect mirror and adjust its distance to the mirror. 

We can extend the domain of applications of our results to situations where we are interested in the scattering of waves for which $k\ell$ is not small, but the inhomogeneity of the slab is confined to a layer of thickness $\delta\ell$ such that $k\delta\ell\ll 1$. This is equivalent to the case where we wish to coat one of the faces of a homogeneous slab of thickness $\ell-\delta\ell$ by a thin layer of inhomogeneous material of thickness $\delta\ell$. Our results yield the low-energy expression for the transfer matrix of the coating $\bM_{\rm c}(k)$. The transfer matrix of the coated slab then follows from the celebrated composition property of transfer matrices \cite{tjp-2020,bookchapter}; it is given by $\bM_{\rm c}(k)\bM_{\rm s}(k)$, where the homogeneous slab and the coating are respectively located in the regions of space determines by $0\leq x\leq\ell-\delta\ell$ and $\ell-\delta\ell\leq x\leq\ell$, and $\bM_{\rm s}(t)$ is the transfer matrix for the homogeneous slab whose explicit form is well-known \cite{ap-2014}. 

\section*{Acknowledgements}
We wish to express our gratitude to Varga Kalantarov for suggesting a couple of references. This work has been supported by the Scientific and Technological Research Council of Turkey (T\"UB\.{I}TAK) in the framework of the project 120F061 and by Turkish Academy of Sciences (T\"UBA).

\ed